\documentclass[%
 sor,
 jor,
 amsmath,amssymb,
preprint,%
author-numerical,%
]{revtex4-2}

\usepackage{amsmath,amssymb,amsfonts}
\usepackage{algorithmic}
\usepackage{hyperref}

\usepackage{graphicx}
\usepackage{textcomp}
\usepackage{textgreek}
\usepackage{braket}
\usepackage{subcaption}
\usepackage{array}
\usepackage{booktabs}
\usepackage{tabularx}
\begin{document}

\title{Transmon qubit modeling and characterization for Dark Matter search}

\author{R. Moretti}
\affiliation{%
Department of Physics - University of Milano-Bicocca - Piazza della Scienza 3 20126 Milan - Italy}
\affiliation{
INFN - Milano Bicocca - Piazza della Scienza - 3 20126 Milan - Italy}%
\affiliation{Bicocca Quantum Technologies (BiQuTe) Centre - 3 20126 Milan - Italy}%

\author{D. Labranca}
\affiliation{%
Department of Physics - University of Milano-Bicocca - Piazza della Scienza 3 20126 Milan - Italy}
\affiliation{
INFN - Milano Bicocca - Piazza della Scienza - 3 20126 Milan - Italy}%
\affiliation{Bicocca Quantum Technologies (BiQuTe) Centre - 3 20126 Milan - Italy}%
\affiliation{RF Technology Division - National Institute of Standards and Technology - Boulder Colorado 80305 - USA}%

\author{P. Campana}
\author{R. Carobene}
\author{M. Gobbo}

\affiliation{%
Department of Physics - University of Milano-Bicocca - Piazza della Scienza 3 20126 Milan - Italy}
\affiliation{
INFN - Milano Bicocca - Piazza della Scienza - 3 20126 Milan - Italy}%
\affiliation{Bicocca Quantum Technologies (BiQuTe) Centre - 3 20126 Milan - Italy}%

\author{M. A. Castellanos-Beltran}
\affiliation{RF Technology Division - National Institute of Standards and Technology - Boulder Colorado 80305 - USA}%

\author{D. Olaya}
\affiliation{RF Technology Division - National Institute of Standards and Technology - Boulder Colorado 80305 - USA}%
\affiliation{Department of Physics - University of Colorado - Boulder Colorado 80309 - USA}

\author{P. Hopkins}
\affiliation{RF Technology Division - National Institute of Standards and Technology - Boulder Colorado 80305 - USA}%

\author {L. Banchi}
\affiliation{Department of Physics and Astronomy - University of Florence - via G. Sansone 1 50019 Sesto Fiorentino (FI) - Italy}
\affiliation{INFN - Firenze - via Sansone 1 50019 Sesto Fiorentino (FI) - Italy}

\author{M. Borghesi}
\affiliation{Department of Physics - University of Milano-Bicocca - Piazza della Scienza 3 20126 Milan - Italy}
\affiliation{
INFN - Milano Bicocca - Piazza della Scienza - 3 20126 Milan - Italy}%
\affiliation{Bicocca Quantum Technologies (BiQuTe) Centre - 3 20126 Milan - Italy}%

\author{A. Candido}
\affiliation{CERN - Theoretical Physics Department - CH-1211 Geneva 23 - Switzerland}

\author{S. Carrazza}
\affiliation{CERN - Theoretical Physics Department - CH-1211 Geneva 23 - Switzerland}
\affiliation{TIF Lab - Department of Physics - University of Milan - Via Celoria 16 20133 Milan - Italy}
\affiliation{INFN - Milano - Via Celoria 16 20133 Milan - Italy}
\affiliation{Quantum Research Center - Technology Innovation Institute - P.O. Box 9639 Abu Dhabi - United Arab Emirates}

\author{H. A. Corti}
\affiliation{Department of Physics - University of Milano-Bicocca - Piazza della Scienza 3 20126 Milan - Italy}
\affiliation{
INFN - Milano Bicocca - Piazza della Scienza - 3 20126 Milan - Italy}%
\affiliation{Bicocca Quantum Technologies (BiQuTe) Centre - 3 20126 Milan - Italy}%

\author{A. D'Elia}
\affiliation{INFN - Laboratori Nazionali di Frascati - 00044 Frascati (RM) - Italy}

\author{M. Faverzani}
\affiliation{Department of Physics - University of Milano-Bicocca - Piazza della Scienza 3 20126 Milan - Italy}
\affiliation{
INFN - Milano Bicocca - Piazza della Scienza - 3 20126 Milan - Italy}%
\affiliation{Bicocca Quantum Technologies (BiQuTe) Centre - 3 20126 Milan - Italy}%

\author{E. Ferri}
\affiliation{
INFN - Milano Bicocca - Piazza della Scienza - 3 20126 Milan - Italy}%
\author{A. Nucciotti}
\author{L. Origo}
\affiliation{Department of Physics - University of Milano-Bicocca - Piazza della Scienza 3 20126 Milan - Italy}
\affiliation{
INFN - Milano Bicocca - Piazza della Scienza - 3 20126 Milan - Italy}%
\affiliation{Bicocca Quantum Technologies (BiQuTe) Centre - 3 20126 Milan - Italy}%

\author{A. Pasquale}
\affiliation{TIF Lab - Department of Physics - University of Milan - Via Celoria 16 20133 Milan - Italy}
\affiliation{INFN - Milano - Via Celoria 16 20133 Milan - Italy}
\affiliation{Quantum Research Center - Technology Innovation Institute - P.O. Box 9639 Abu Dhabi - United Arab Emirates}

\author{A. S. Piedjou Komnang}
\author{A. Rettaroli}
\author{S. Tocci}
\affiliation{INFN - Laboratori Nazionali di Frascati - 00044 Frascati (RM) - Italy}

\author{C. Gatti}
\affiliation{INFN - Laboratori Nazionali di Frascati - 00044 Frascati (RM) - Italy}

\author{A. Giachero}
\affiliation{Department of Physics - University of Milano-Bicocca - Piazza della Scienza 3 20126 Milan - Italy}
\affiliation{
INFN - Milano Bicocca - Piazza della Scienza - 3 20126 Milan - Italy}%
\affiliation{Bicocca Quantum Technologies (BiQuTe) Centre - 3 20126 Milan - Italy}%
\affiliation{Quantum Sensors Division - National Institute of Standards and Technology - Boulder Colorado 80305 - USA}

\begin{abstract}

This study presents the design, simulation, and experimental characterization of a superconducting transmon qubit circuit prototype for potential applications in dark matter detection experiments. We describe a planar circuit design featuring two non-interacting transmon qubits, one with fixed frequency and the other flux tunable. Finite-element simulations were employed to extract key Hamiltonian parameters and optimize component geometries. The qubit was fabricated and then characterized at $20$ mK, allowing for a comparison between simulated and measured qubit parameters. Good agreement was found for transition frequencies and anharmonicities (within 1\% and 10\% respectively) while coupling strengths exhibited larger discrepancies (30\%). We discuss potential causes for measured coherence times falling below expectations ($T_1\sim\,$1-2 \textmu s) and propose strategies for future design improvements. Notably, we demonstrate the application of a hybrid 3D-2D simulation approach for energy participation ratio evaluation, yielding a more accurate estimation of dielectric losses. This work represents an important first step in developing planar Quantum Non-Demolition (QND) single-photon counters for dark matter searches, particularly for axion and dark photon detection schemes.
\end{abstract}

\maketitle

\section{Introduction}
\label{sec:introduction}
Superconducting qubits have emerged as leading candidates for a wealth of quantum sensing applications \cite{schuster,qsensingrev, danilin2024quantum}, owing to their coherence-preserving properties and excellent sensitivity to microwave photons. In the last two decades, significant advancements have been demonstrated in detecting and controlling individual quanta, such as photons \cite{schuster}, phonons \cite{o’connel____2010,2015,Qacustics,Qbar,qubitbaw, remoteentangl}, and magnons \cite{magn1, magn2}. Progress in quantum sensing is closely tied to developments in quantum computer engineering, which has propelled us into the era of quantum utility \cite{util1, util2}, driven by the collective efforts of private companies, research institutions, and universities.

Specific qubit implementations, including transmon, flux qubit, and fluxonium \cite{koch, orlando, manucharyan}, find applications in fundamental physics experiments, such as the search for weakly electromagnetic (EM) coupled dark matter (DM) candidates like axions \cite{tomonori, braggio} and dark photons \cite{dixit, dixit2}.

The qubit platforms mentioned above benefit from extended coherence times and high sensitivity to AC fields \cite{najerasantos}. As such, they can be effectively driven by photons resulting from DM-EM interactions. Novel detection schemes have been developed, such as the Quantum Non-Demolition (QND) technique \cite{qnd, dixit} and Direct Detection (DD) \cite{chen} through qubit excitations. QND leverages the non-adiabatic interaction between photons trapped in a cavity and a dispersively coupled qubit. The interaction induces an AC-Stark effect such that the Jaynes-Cummings Hamiltonian $\mathcal{H}$ \cite{Jaynes:1963zz} can be modeled as:
\begin{equation}
\mathcal{H}= \omega_r a^\dag a + \frac{1}{2}\left(\omega_q + 2\xi a^\dag a \right)\sigma_z,
\end{equation}
$\omega_r$ and $\omega_q$ are the storage cavity and qubit's first transition frequency, $a^\dag$ and $a$ are the cavity creation and annihilation operators, respectively. The term $\xi$ is a qubit frequency shift depending on its coupling strength with the storage cavity.
Hence the qubit initially prepared in state ${\ket{\psi_0}=\left(\ket{0} + \ket{1}\right) / \sqrt{2}}$ acquires a precession term:
\begin{equation}
\label{eq:precession}
\ket{\psi(t)} = \frac{1}{\sqrt{2}} \left(\ket{0} + e^{-2in\xi t}\ket{1}\right),
\end{equation}
$n=a^\dag a$ is the number of photons that populate the storage cavity.
This phase term is time-dependent and proportional to the number of photons trapped in the cavity. This allows for the photon number to be inferred through parity measurements of the qubit state. Since this type of detection does not destroy photons (hence the term "non-demolition"), it is possible to repeat the inference procedure several times, reducing dark counts, and effectively mitigating inefficiencies caused by readout errors.

This detection method has already been demonstrated for dark photon searches \cite{dixit}. However, axion detection poses an additional challenge, requiring a strong magnetic field inside the storage cavity to enable photon conversion via the Inverse Primakoff effect \cite{primakoff, primakoff2}.
Therefore, the storage cavity and the qubit must be positioned at a considerable distance from the field. This challenge can be overcome by implementing itinerant photon detection techniques \cite{itinerant1, itinerant2}. An extreme solution involves achieving remote entanglement between qubits placed in different cryostats, as demonstrated in \cite{magnard}. Alternatively, a qubit can be installed in the same refrigerator as a haloscope (i.e.\ a tunable microwave cavity with high-quality factor immersed in a strong magnetic field \cite{originalhaloscope}) and connected via a coaxial cable \cite{braggio}.

\begin{figure}
    \centering
    \includegraphics[width=0.8\linewidth]{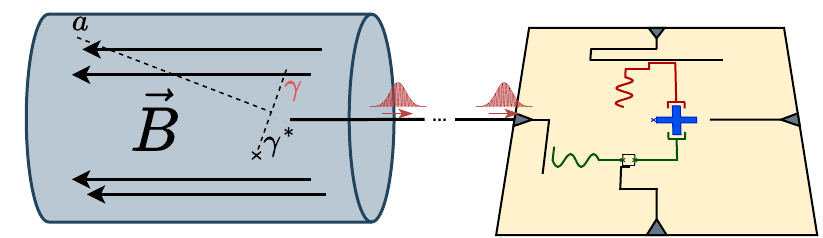}
    \caption{Schematic representation of a QND axion detection scheme. The axion field $a$, interacting with the magnetic field $\vec{B}$ inside the haloscope, produces a photon ($\gamma$) and a virtual photon ($\gamma^*$). The photon is absorbed by a tunable storage cavity (green), inducing a phase precession in the qubit (blue) state. Repeated photon detection is achieved by performing parity measurements via the readout resonator (red).}
    \label{fig:detection_setup}
\end{figure}

Our proposal addresses the challenge of introducing a magnetic field into the experimental setup, similar to the approach described in \cite{braggio}, while incorporating modifications to enable quantum non-demolition (QND) detection. In the method presented in \cite{braggio}, photons produced in the haloscope via the axion conversion process exit the haloscope and propagate toward a photon counter. This system employs a frequency-tunable cavity that efficiently absorbs resonant photons, leading to qubit excitation via a four-wave mixing process enabled by a dedicated pump line. In contrast to this detection mechanism, which is destructive, we propose replacing the aforementioned photon counter with a QND-compatible detector, in analogy with the one developed in \cite{dixit}. Although the setup in \cite{dixit} employs a three-dimensional (3D) storage cavity, in our work two-dimensional (2D) cavities are explored.
This alternative would offer a more compact, fully on-chip implementation, despite reducing the quality factor. Notably, the proposed device shares the same architecture as that used in the direct excitation method; however, certain lines are repurposed, as the four-wave mixing process is no longer required. The system consists of a transmon qubit coupled to a readout resonator, a high-quality tunable storage cavity, and a dedicated driveline. Photons originating from the haloscope and traveling along the transmission line are trapped within the storage cavity when resonant, with a lifetime $\tau = Q_s/\omega_s$, where $Q_s$ is the storage cavity quality factor and $\omega_s$ the resonant frequency. During this time, the qubit state undergoes precession according to Eq. \ref{eq:precession}. Provided that $\tau$ is significantly longer than the qubit readout time, multiple non-destructive measurements can be carried out, achieving dark count suppression. Fig. \ref{fig:detection_setup} illustrates the proposed detection scheme.

From a theoretical perspective, QND photon counters can benefit from the interaction between a storage cavity and multiple qubits to enhance dark count suppression and eliminate uncorrelated noise effects, thereby requiring fewer measurements to reach the same sensitivity and accelerating the experiment. Additionally, multi-qubit quantum states, such as the Greenberger–Horne–Zeilinger (GHZ) state, can introduce different --and possibly advantageous-- measurable precession terms. For instance, the precession term for a GHZ state becomes \cite{degen}:
\begin{equation}
\ket{\psi(t)} = \frac{1}{\sqrt{2^m}} \left(\ket{00...0} + e^{-2inm\xi t}\ket{11...1}\right),
\end{equation}
the enhancement term $m$ corresponds to the number of qubits composing the state. 

While this study will not explore the Direct Detection scheme in detail, we highlight its potential due to the simplicity of its experimental implementation and scalability \cite{chen}. A flux-tunable qubit can be used to detect low-power coherent AC drives when in resonance with a qubit transition frequency, making this setup suitable for detecting dark photons due to their kinetic mixing with the EM field \cite{BRAHMACHARI2014441}. This drive induces a comparably slow Rabi oscillation of the qubit state, that can be predicted analytically as a function of the kinetic mixing parameter $\epsilon$. Hence, if a qubit is prepared in the ground state \(|0\rangle\), it will accumulate an excitation probability over time that can be estimated through repeated measurements. A spike in the excitation probability occurs when the qubit resonant frequency matches the dark-photon field-induced drive.

Both detection mechanisms require high-performance qubits in terms of trade-off between coherence time and readout speed. This study, consisting of a follow-up of \cite{moretti}, presents our design of a test superconducting qubit, comparing simulation results with experimental measurements of parameters such as coupling strengths and coherence times. We assess the strengths and limitations of these simulation techniques as we move toward developing a QND photon counter.

\section{Circuit design and simulation}
\label{sec:sim}
In this section, we present a planar circuit design featuring two non-interacting transmon qubits. This test circuit enables evaluation of our design and modeling capabilities before advancing to the engineering of the photon-collection cavity and detection setups.

\begin{figure}[h]
    \centering
    \includegraphics[width =1\linewidth]{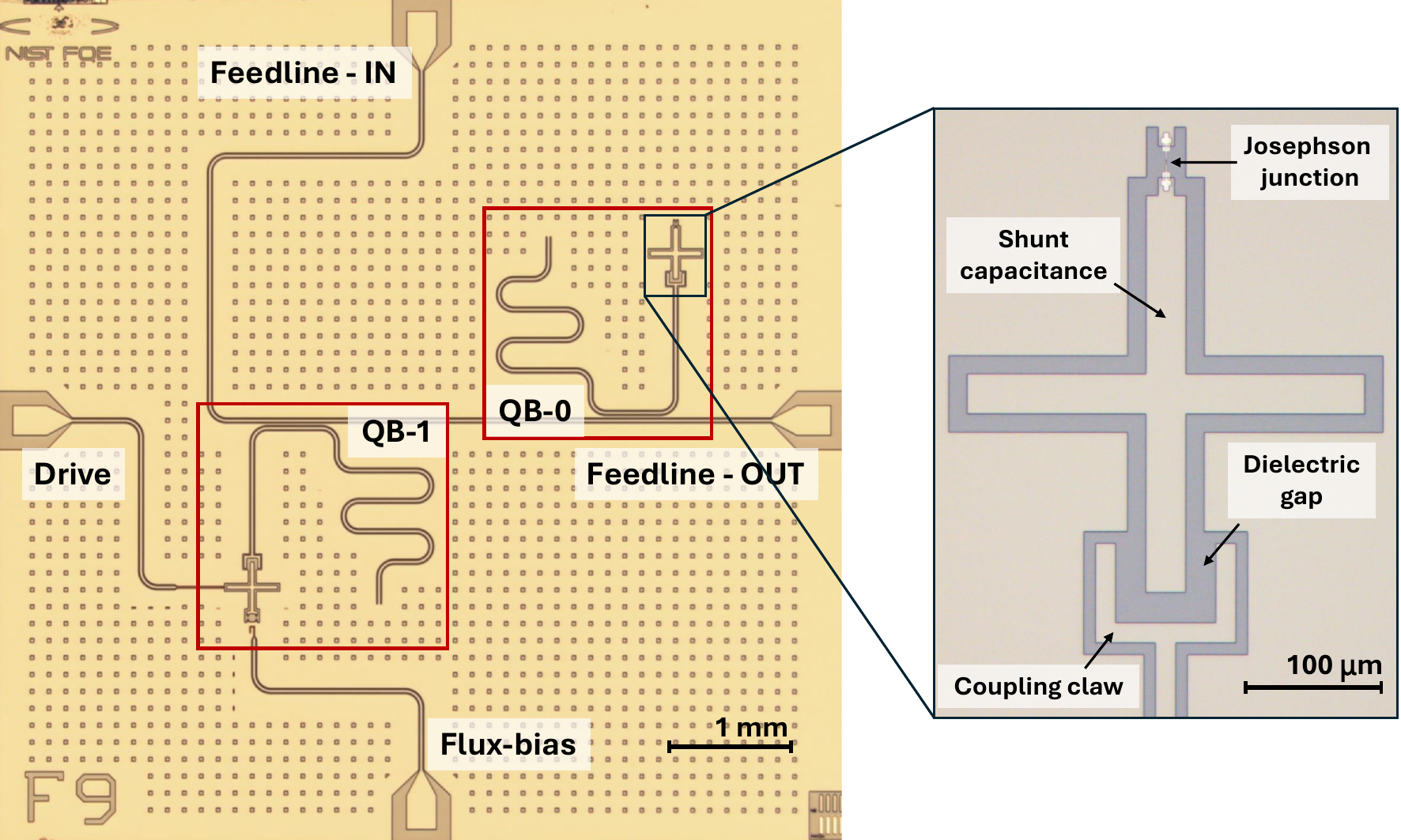}
    \caption{Micrograph of the $7\times7$ mm${}^2$ circuit. Labels indicate the feedline ends for both qubits, the driveline, and the flux-bias line for the flux-tunable qubit. The fixed-frequency qubit is labeled as 'QB-0', and the tunable-frequency qubit as 'QB-1'. The red boxes enclose the QB-0 and QB-1 qubit-resonator systems, with their feedline couplings. On the right, a zoomed-in view of QB-0 shows a detailed view of the qubit components and geometry.}
    \label{fig:circuit}
\end{figure}

The circuit is fabricated on a $7\times 7$ mm${}^2$ high-resistivity silicon substrate of $380$ \textmu m thickness. The qubits are made of a $100$ nm-thick Nb layer with Al/AlOx/Al Dolan-bridge Josephson Junctions \cite{dolan}. The two qubits are grounded transmons shunted with an "x"-shaped capacitance (Xmon \cite{xmon}) and capacitively coupled to the same feedline through individual $\lambda / 4$ resonators. The first qubit has a fixed frequency and is driven through the resonator, while the second qubit is flux-tunable, with a dedicated drive line and flux bias line. Fig. \ref{fig:circuit} shows a micrograph of the circuit. The design was developed using IBM's Qiskit Metal toolkit \cite{qiskitmetal} for qubit circuit prototyping and simulated with various finite-element solvers such as Ansys High-Frequency Structure Simulator (HFSS), Ansys Q3D, and Elmer FEM Solver\footnotemark[1]\footnotetext[1]{Certain equipment, instruments, software, or materials, commercial or non-commercial, are identified in this paper in order to specify the experimental procedure adequately. Such identification does not imply recommendation or endorsement of any product or service by NIST, nor does it imply that the materials or equipment identified are necessarily the best available for the purpose.}.

Simulation is a crucial step as it allows us to extract several Hamiltonian parameters and cross-Kerr terms, which must be carefully optimized by tuning component geometries such as the shunt capacitor, dielectric gaps, and coupler dimensions.

The Xmon cross has a total length of $300$ \textmu m, with a gap of $14$ \textmu m from the ground plane. The cross width for QB-0 measures $30$ \textmu m and $21$ \textmu m for QB-1. These dimensions ensure a total qubit capacitance $C_\Sigma = 100$ fF (QB-0) and $C_\Sigma = 93$ fF (QB-1), without considering the Josephson junction (or SQUID) capacitance and the capacitive load due to resonator coupling. These configurations allow us to have a reasonable first qubit frequency (approximately between $4$ GHz and $6$ GHz) and fall in the transmon regime for a wide range of Josephson inductance values $L_J \in [7\,\text{nH},\,15\, \text{nH}]$ \cite{qeg, manentimotta}. This regime is defined by an inductive energy $E_J$ to capacitive energy $E_C$ ratio $E_J / E_C > 50$.

The coplanar waveguide (CPW) readout resonators are about $3.68$ mm and $3.77$ mm long for the fixed-frequency and tunable frequency qubits respectively, with a trace width of 15 \textmu m and a 9 \textmu m gap, resulting in a characteristic impedance of $Z_0= 50$ \textOmega, assuming a relative dielectric permittivity of $\epsilon_r=11.65$ \cite{epsr}. The coupling element is a capacitive claw with the same trace width and gap as the resonator, and $23$ \textmu m distant from the metal cross. Additionally, the resonator is capacitively coupled to the feedline through a $330$ \textmu m long coupling segment at a $30$ \textmu m distance from the feedline.

Relevant couplings were derived from the capacitance matrices extracted with Ansys Q3D for each qubit-readout coupler subsystem, assuming a junction capacitance of $2$ fF. This derivation was automatically performed using the lumped oscillator model (LOM) \cite{lom}, calculated as a function of the Josephson inductance $L_J$. A more accurate evaluation can be performed with Ansys HFSS Eigenmode by extracting the resonant frequencies accounting for the fully distributed qubit-resonator subsystems. This method also allows us to carry out an energy participation ratio (EPR) \cite{epr} analysis, which quantifies how much energy of a mode is stored in each element to evaluate several parameters of interest, such as dielectric losses and Kerr coefficients. For a single nonlinear element (either single junctions or SQUIDs):
\begin{equation}
\chi_{nm} = \frac{\hbar\,\omega_m\,\omega_n}{4E_{\text{J}}}\,p_m\, p_n,
\end{equation}
where the anharmonicities are $\alpha_m = \chi_{mm}/2$ and the total dispersive shifts $\chi_{nm}$ for $n \neq m$. Here, $p_m$ and $p_n$ refer to the energy participation ratios of the Josephson element for modes $m$ and $n$. For simplicity, we refer to $\chi$ as the resonator's total dispersive shift induced by qubit transitions $\ket{g} \leftrightarrow \ket{e}$, and to $\alpha$ as the qubit anharmonicity.

\begin{figure}
    \centering
    \includegraphics[width=0.8\linewidth]{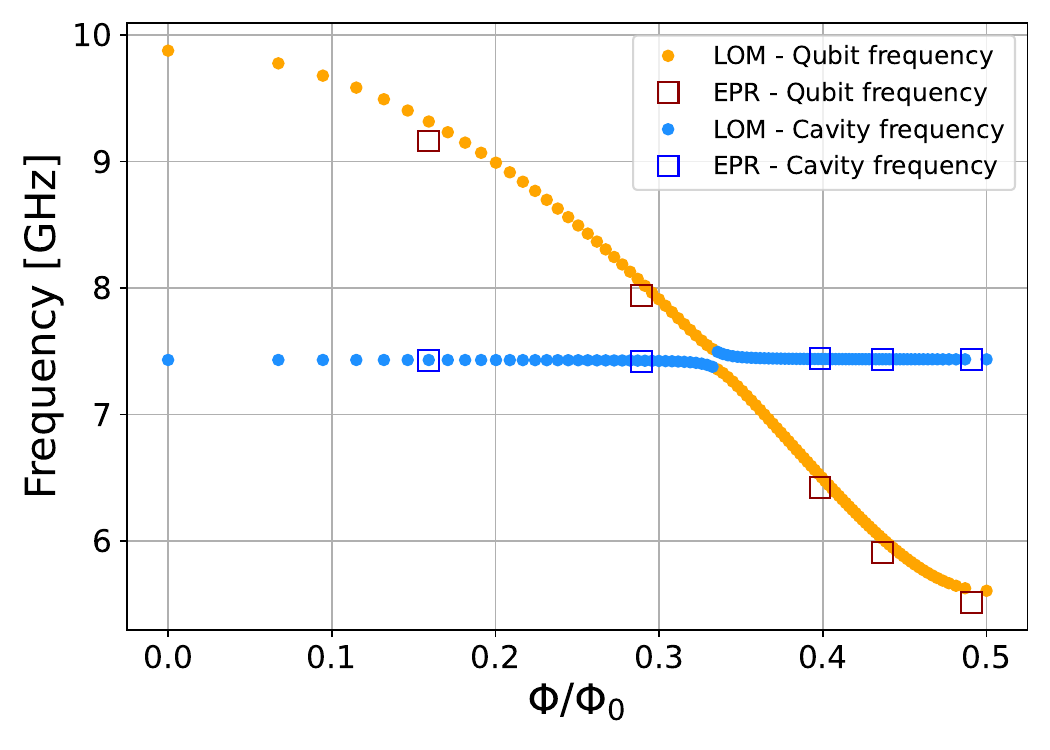}
    \caption{Comparison between EPR and LOM frequency estimations for our design at different flux bias conditions.}
    \label{fig:frequencies}
\end{figure}

It is important to note that Ansys HFSS does not allow finite-element simulations of nonlinear inductances. Therefore, we adopt a linear approximation, which is sufficient for weakly anharmonic qubits such as the transmon. The frequency of the mode $\omega_{q\text{lin}}$ corresponding to the qubit ${\ket{g} \rightarrow \ket{e}}$ transition is corrected analytically in a later stage of the EPR analysis \cite{epr}. The results obtained with the LOM and the EPR methods are consistent and align with expectations \cite{lom,epr,moretti}, as exemplified in Fig. \ref{fig:frequencies} in which we compare the resonant frequencies of a resonator-tunable qubit system for different flux biases. The LOM analysis requires the dressed resonator frequency $\omega_r$ as an input, so we used the value extracted through the EPR method.
We also conducted a study of dielectric losses through EPR analysis. We modeled the dielectric quality factor $Q_{\text{TLS}}$ contributions by decomposing the different material contributions:
\begin{equation}
\frac{1}{Q_{\text{TLS}}} = \sum_{i}^{}p_i \tan{\delta_i},
\end{equation}
where $p_i$ and $\tan{\delta_i}$ are the participation ratios of the material and geometry-dependent loss tangents, respectively. These ratios are determined by the fraction of energy stored within a volume enclosing the Josephson element relative to the total energy stored in the qubit-resonator system. Estimating the participation ratio in thin layers, such as the oxide regrowth between material interfaces, requires accurate modeling of the electric field near metal edges. This is challenging to achieve with three-dimensional simulations because they require extremely fine-grained meshing in the region of interest, often making such simulations too inefficient for standard desktop computing. This problem can be addressed using analytical models \cite{Martinis2022, Eun2023} or a hybrid 3D-2D simulation strategy \cite{wang}.

To estimate dielectric losses for our qubit design, we partitioned the three-dimensional layout into regions of interest to estimate the total stored energy with Ansys HFSS. We then assigned participation ratios to the relevant materials for the regions of interest based on values extracted from two-dimensional (cross-section) simulations performed with Elmer FEM Solver. We carried out this procedure by replicating the Xmon design with KQCircuits\footnotemark[1] \cite{kqc}, a circuit design tool developed at IQM Quantum Computers that allows us to automatically implement the hybrid method. The reduced complexity of the 2D simulation allows for fine-grained meshing, capturing the field distribution around metal edges accurately. By combining it with the comprehensive modeling of 3D layouts, we achieve a more accurate estimation of surface participation ratios without the computational burden of extremely fine 3D meshing. This works under the assumption that the field distribution across different materials does not change significantly along the region of interest, and only the overall field intensity varies. The approximation becomes less accurate when the same partition region embeds gaps of different sizes. We can, however, mitigate such an error by introducing multiple partition regions and associating them with dedicated cross-section simulations. While this particular feature is available within KQCircuits, its application is still not widely documented in the literature. By showcasing this analysis, we aim to highlight its potential value for future qubit designs, particularly in applications requiring loss estimations.

\begin{figure}[ht]
    \centering
    \begin{subfigure}[b]{\linewidth}
        \centering
        \includegraphics[width=0.8\linewidth]{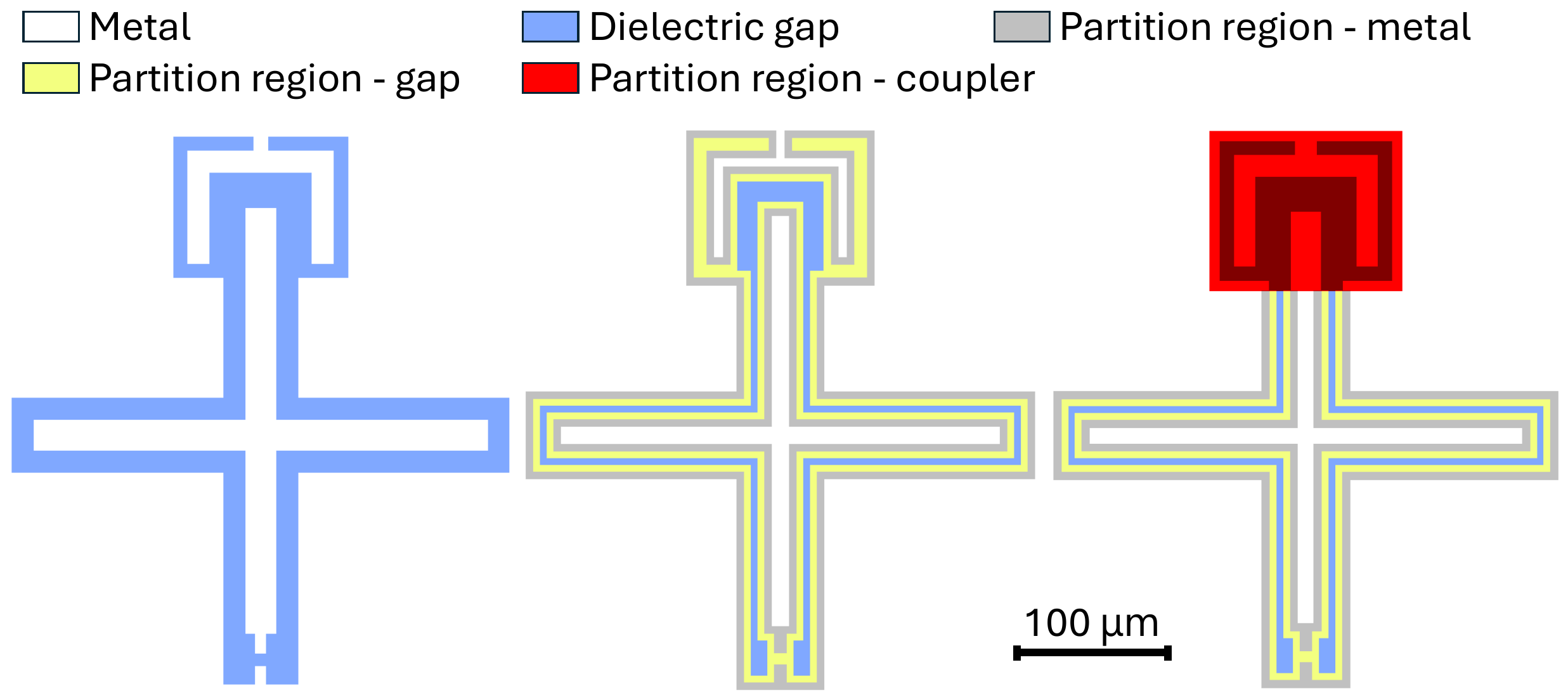}
        \caption{}
        \label{fig:partitions_a}
    \end{subfigure}
    \vspace{1em} 
    \begin{subfigure}[b]{0.8\linewidth}
        \centering
        \includegraphics[width=\linewidth]{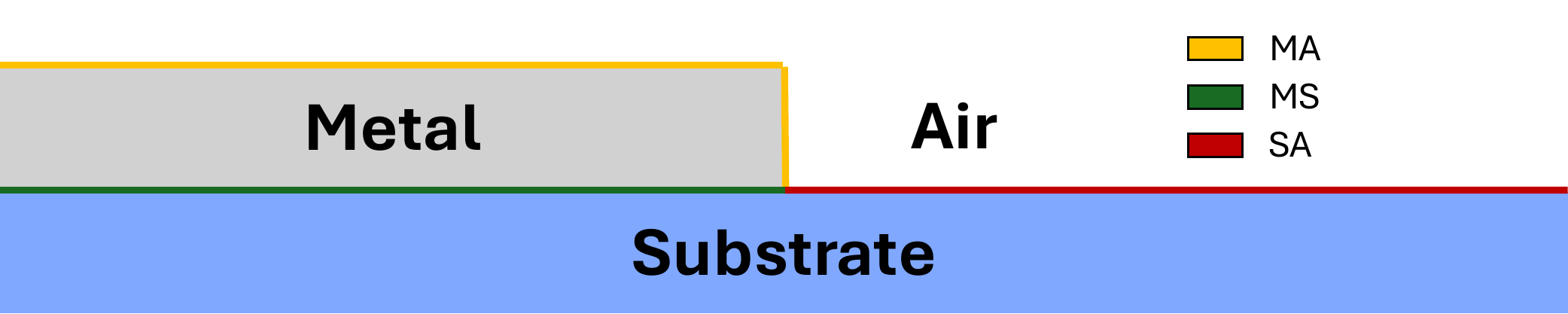}
        \caption{}
        \label{fig:partitions_b}
    \end{subfigure}
    \caption{Graphic overview of the hybrid 3D-2D simulation for dielectric losses estimation. (a) Three simulation strategies. No partition (left), partition along the metal-edge region, extending $5$ \textmu m towards metal and $5$ \textmu m inside the gap area (center) and partition along the metal-edge region and coupler region (right). To each participation region corresponds a transverse cut and a dedicated cross-section simulation. (b) Cross section with highlighted interfaces: metal (M), substrate (S) and air (A). Thicknesses are not to scale. Fine-grained meshing for these submicrometric structures is easily achievable in two dimensions.}
    \label{fig:partitions}
\end{figure}

Fig. \ref{fig:partitions} illustrates the different partitioning approaches that we considered. In the first approach, we relied on the three-dimensional simulation. We next partitioned the metal-edge region, leveraging the fact that the cross-gap is uniform everywhere, excluding the areas in proximity to the coupler and the junction wire. Finally, we introduced an additional partition region containing the coupling claw to further correct the surface EPR estimation.
Table \ref{tab:epr} collects the surface EPRs for the three approaches depicted in Fig.\ref{fig:partitions_a}. It is evident that three-dimensional simulation considerably underestimates surface EPR, which would directly lead to an incorrect estimation of $T_1^{\text{TLS}}$. As expected, the difference is particularly pronounced for the MS surface, which is the thinnest one, while the two hybrid approaches are roughly equivalent, differing at the percent level. We note that an additional partition region for the junction wire could have also been considered \cite{Martinis2022}, but this would have been less precise due to the absence of an accurate junction model for our qubits.

\begin{table}[ht]
    \centering
    \caption{Surface EPR values estimated through 3D-only or hybrid 3D-2D approaches. The thicknesses $\sigma$ for each interface are assigned as measured in \cite{thick} for niobium-on-silicon CPW  resonators.}
        \begin{tabular}{cccc}
            \toprule
             & 3D only & 3D-2D (MER) & 3D-2D (MER-coupler) \\
            \midrule
            MA & $6.3\times 10^{-6}$& $8.93\times 10^{-5}$ & $8.84\times 10^{-5}$ \\
            MS & $3.06\times 10^{-7}$& $4.80\times 10^{-5}$ & $4.75\times 10^{-5}$ \\
            SA & $1.17\times 10^{-4}$ & $1.59\times 10^{-4}$ & $1.58\times 10^{-4}$ \\
            \bottomrule
            \multicolumn{4}{l}{$\sigma_{\text{MA}}=4.8$ nm $\;\;\;\;\;\;\sigma_{\text{MS}}=0.3$ nm $ \;\;\;\; \;\;\sigma_{\text{SA}}=2.3$ nm}
        \end{tabular}
    \label{tab:epr}
\end{table}
By considering loss tangent values as indicated in \cite{losst}, we can derive the internal dielectric quality factor $Q_{\text{TLS}}$ for the qubit. Utilizing the hybrid approach, with a dedicated partition region for the coupler, we obtain $Q_{\text{TLS}}=7.81\times 10^5$, as compared to the $Q_{\text{TLS}}=2.62\times 10^6$ value estimated from the 3D approach.
\section{Experimental characterization}
\label{sec:experiment}
\begin{figure}
    \centering
    \includegraphics[width=\linewidth]{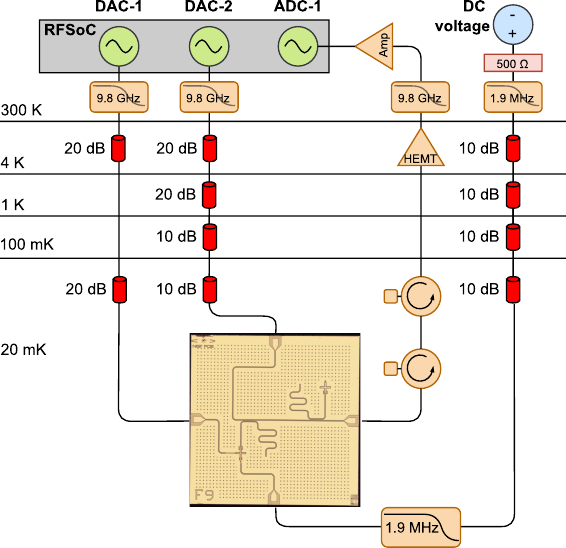}
    \caption{Experimental setup used for the test circuit characterization.}
    \label{fig:setup}
\end{figure}

The qubit device was fabricated at the US National Institute of Standards and Technology (NIST) and measured at the cryogenic facilities of the University of Milano-Bicocca. The experimental setup is depicted in Fig. \ref{fig:setup}. The qubit was placed inside a dilution refrigerator at a temperature of $20$ mK. A voltage generator produced the DC signal for flux bias, while three RF lines (feedline input-output and driveline input) were generated by a single RFSoC4x2 board, with a DAC sampling rate of $9.85$ GSPS and an ADC sampling rate of $5$ GSPS. Attenuations were distributed across different temperature stages, for a total of $-40$ dB on the driveline and flux-bias line, and $-60$ dB on the feedline input. The first-stage amplification of the feedline output was given by a low-noise High Electron Mobility Transistor (HEMT) amplifier (Low Noise Factory part number LNF-LNC4\_8C) at $4$ K, providing a $40$ dB gain. Two circulators (QuinStar and LNF-CIC4\_12A $4$-$12$ GHz) were positioned between the qubit and the HEMT, achieving a $-40$ dB noise suppression from the HEMT to the qubit. Additional amplification of $26.5$ dB was added at room temperature by means of a Mini-Circuits ZX10-2-183-S+ amplifier. Two low-pass filters (Mini-Circuits) with a $1.9$ MHz cutoff were applied to the DC line, one at room temperature and one at the mixing chamber, and three low-pass filters (Mini-Circuits 16642 ZXLF-K982+) at $9.8$ GHz were applied at the DAC and ADC lines output to suppress spurious harmonics generated by the RFSoC\footnotemark[1].

\begin{figure}
    \centering
    \includegraphics[width=0.8\linewidth]{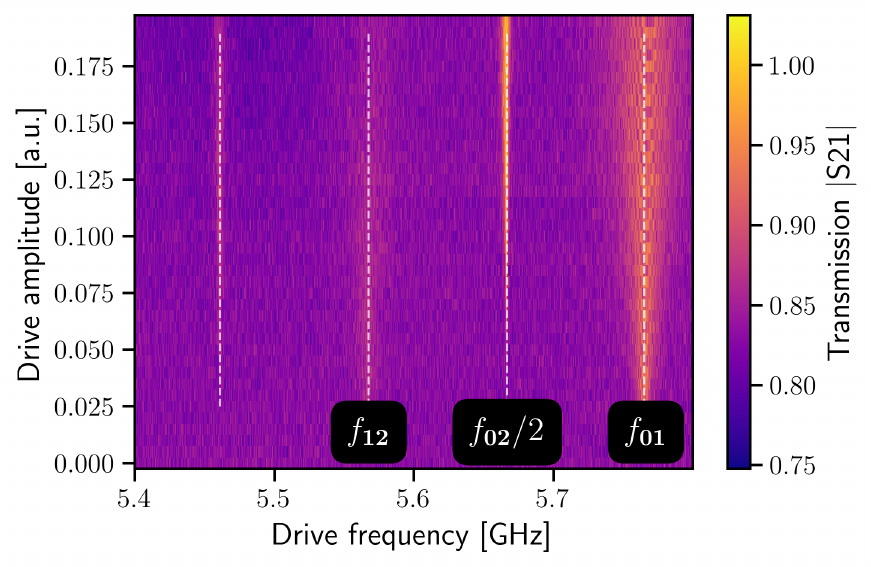}
    \caption{Two-tone spectroscopy of the fixed frequency qubit, where the readout signal transmission intensity was registered for different drive frequencies and power.}
    \label{fig:twotone}
\end{figure}

We controlled the RF signals through the Qibolab software\footnotemark[1]~\cite{Efthymiou2024,Carobene2023,Pasquale2024}, which provided complete control over RF signal generation and the realization of calibration experiments. In this study, we report the fundamental measurement outcomes for both qubits. Fig. \ref{fig:twotone} shows the qubit spectroscopy of the fixed-frequency qubit, revealing the transition frequencies up to the third-excited state. The measurement utilized two-tone spectroscopy, where the readout and drive tones were sent continuously to the qubit while varying the drive frequency and power. When the drive frequency is resonant with a qubit transition, the readout resonator frequency shifts, producing a measurable change in the readout signal transmission in-phase (I) and quadrature (Q) components.

\begin{figure}
    \centering
    \includegraphics[width=0.8\linewidth]{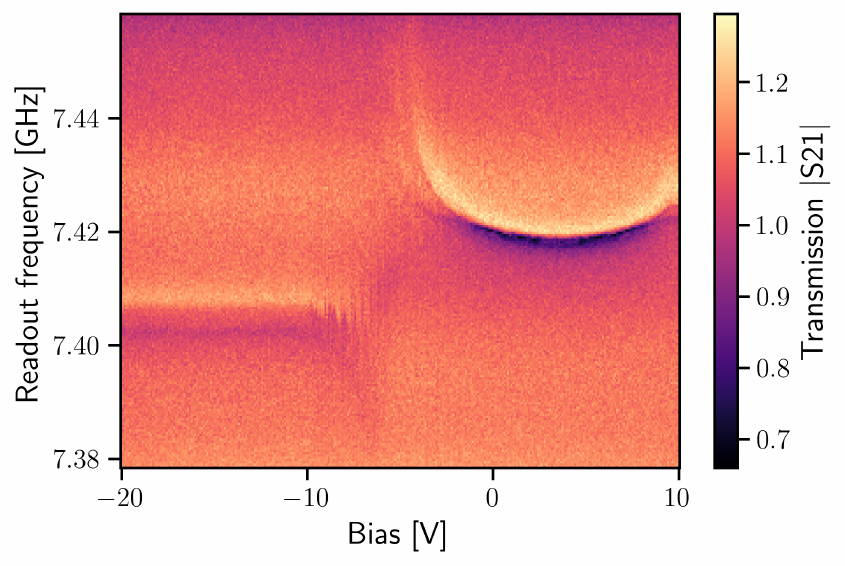}
    \caption{Flux-tunable qubit resonator spectroscopy as a function of the flux bias.}
    \label{fig:avoided_crossing}
\end{figure}

\begin{figure}
    \centering
    \includegraphics[width=0.8\linewidth]{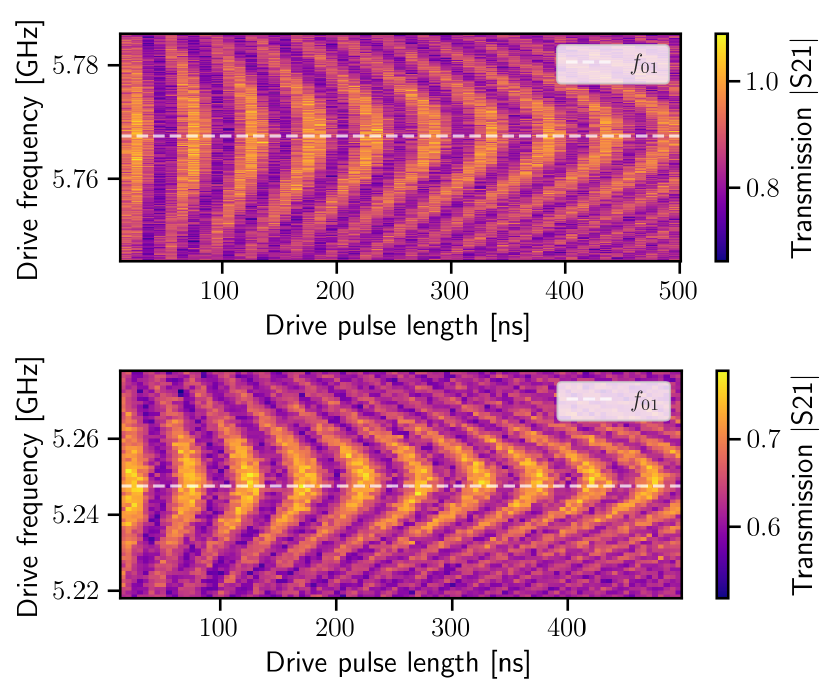}
    \caption{Chevron plot of the fixed-frequency (top) and flux-tunable (bottom) qubits, showing the Rabi oscillations for the transition $\ket{0}\leftrightarrow\ket{1}$.}
    \label{fig:chevron}
\end{figure}

For the flux tunable qubit, the resonator spectroscopy was carried out as a function of the DC voltage at the source, which is linearly related to the applied flux bias to the SQUID. The results are depicted in Fig. \ref{fig:avoided_crossing} and exhibit avoided crossing \cite{bishop2009nonlinear} when the qubit is in resonance with its readout resonator.
The coherence times have also been measured. Fig. \ref{fig:chevron} presents the Chevron plots of the fixed-frequency and tunable qubits, showing oscillations that decay in approximately $1$ \textmu s. For the fixed (tunable) frequency qubit, the experiment has been carried out by sending the drive pulse through the feedline (driveline) and tuning the pulse power to reach a similar Rabi frequency. Compared to driving through the feedline, driving the qubit through the driveline allowed us to send approximately $12$ dB less in power to achieve the same Rabi frequency.
Table \ref{tab:results} summarizes the characterization results and compares them with expected values extracted from simulations. The latter are set \textit{a posteriori}, by tuning the $L_J$ input in the LOM to match the experimentally measured frequencies, then retrieving all the other parameters through an Ansys HFSS Eigenmode simulation and the corresponding EPR quantization. The results for the flux tunable qubit (QB-1) refer to a zero-flux bias working point. The table also includes measurements and expected values of the readout resonator internal ($Q_i$) and coupling ($Q_c$) quality factors.

\begin{table}[ht]
    \centering
    \caption{Qubit characterization results, showing the agreement between measured and expected values. $T^{*}_2$ and $T_2$ refer to Ramsey and Hahn-echo characterization experiments \cite{qeg}, respectively.}
    \begin{tabularx}{\columnwidth}{lXXXX}
        \toprule
         & QB-0 & QB-1 & Exp. QB-0 & Exp. QB-1\\
        \midrule
        $\omega_{q}/2\pi$ [GHz] & $5.7660(5)$ & $5.2483(5)$ & $5.6941$ & $5.1990$ \\
        $\alpha /2\pi$ [MHz] & $-198.6(14)$ & $-185.0(14)$ & $-195.3$ & $-199.1$\\
        $g / 2\pi$ [MHz] & $80(11)$ & $68(6)$ & $115$ & $105$\\
        $\chi / 2\pi$ [kHz]& $350(98)$& $168(28)$ & $646$ & $398$\\
        $\omega_r / 2\pi$ [GHz] & $7.57905(4)$ & $7.419143(6)$ & $7.60728$ & $7.448263$\\
        $Q_i / 10^3$ & $15.3(16)$ & $7.62(8)$ & $> 10^2$ & $> 10^2$\\
        $Q_c / 10^3$ & $4.28(8)$ & $7.30(4)$& $10.9$& $10.9$\\
        $T_1$ [\textmu s] & $1.52(5)$ & $1.93(14)$ & $\lesssim 12$ \textmu s & $\lesssim 18$ \textmu s\\
        $T^{*}_2$ [\textmu s] & $0.229(7)$ & $0.586(53)$ & - & - \\
        $T_2$ [\textmu s] & $0.61(3)$ & $1.03(10)$ & $\lesssim 2T_{1}$ & $\lesssim 2T_{1}$\\
        \bottomrule
    \end{tabularx}
    \label{tab:results}
\end{table}

Our method accurately predicts the qubit frequencies and anharmonicities (${\sim 1\%}$ for $\omega_q$, $\omega_r$, and $\sim 10\%$ for $\alpha$). Other Hamiltonian parameter estimations, i.e.\ the coupling strengths $g$, are overall less accurate ($\sim 30\%$ higher). This reflects an even greater discrepancy with the dispersive shifts $\chi$, due to the quadratic dependence on $g$ \cite{qeg, manentimotta}. However, we must take into account the difficulty of estimating $\chi$ experimentally, mainly due to the reduced coherence time, resulting in a distorted resonance shape when the qubit is in the excited state. Also, the prediction of coupling quality factors $Q_c$ only seems accurate up to an order of magnitude. This might be affected by the presence of standing waves and impedance mismatches that are hard to simulate (and possibly due to the measurement setup) that have a detrimental effect on the quality factors \cite{besedin2018quality}.

The coherence times of both qubits fall significantly below expectations. The experimental values for $T_1$, $T_2$, and $T_2^*$, reported in Table \ref{tab:results} are the average results of several individual measurements taken over around $50$ hours. The expected values for $T_1$ account for dielectric losses and Purcell decay through the resonator. The quality factors considered for this estimate include the dielectric qubit quality factor $Q_{\text{TLS}}$ estimated in Sec. \ref{sec:sim} and a resonator quality factor estimated from measured internal and coupling quality factors $Q_r = (1/Q_i + 1/Q_c)^{-1}$. We note that using the 3D-2D simulation method described in Sec. \ref{sec:sim}, rather than the fully 3D approach, adjusts $Q_{\text{TLS}}$ to a lower value, reducing $T_{1\text{TLS}} = Q_\text{TLS}/\omega_q$ from approximately $75$ \textmu s to about $22~\text{\textmu s}$. This decrease is qualitatively consistent with the low $T_1$ measured experimentally. However, this result alone is insufficient to demonstrate the advantage of the hybrid simulation conclusively. To strengthen this assessment, measurements on additional qubits are needed to ensure reproducibility. A further improvement would include replacing loss tangent values reported in the literature with experimentally measured ones specific to the same fabrication process as the measured quantum devices in use.

While the exact cause of the limited coherence times has yet to be fully determined experimentally, we attribute this at least partially to non-idealities in the fabrication process. This hypothesis is supported by the measured internal quality factor of the readout resonator, which is significantly lower than typical target values for transmon qubit readout. Additional factors affecting coherence times include the overall measurement setup, such as the presence of spurious modes, or non-ideal qubit thermalization. The short dephasing time ($T_2 < T_1$) also suggests these issues. 
From a design perspective, improvements can be made by reducing the surface participation ratio. This can be achieved, in principle, by increasing the overall dimensions of the Xmon (gaps and metal cross) to the point where dielectric loss is primarily limited by the substrate. Other directions involve the investigation of different qubit geometries, such as the floating double-pad shape that typically exhibits high coherence \cite{kono2023, wang2022} and experimenting with novel fabrication strategies such as surface encapsulation \cite{bal2024systematic}.

\section{Conclusions}
The design, simulation, and experimental characterization of the transmon qubit presented in this study have yielded significant insights into qubit design and the retrieval of parameters of interest through simulations and experiments. These parameters include coupling capacitances, coupling strengths, and transition frequencies. This initial step was fundamental in assessing the degree of control over these parameters before developing planar Quantum Non-Demolition (QND) single-photon counter circuits. Our usage of the hybrid 3D-2D simulation approach for EPR evaluation, as implemented in KQCircuits, demonstrates a novel, yet underutilized method for estimating surface participation ratios and dielectric losses in superconducting qubit designs. This approach could prove beneficial for future studies in this field.

Experimental characterization at cryogenic temperatures demonstrated the coherent control of the fabricated device, allowing us to assess the degree of agreement with simulations carried out with the help of finite-element analyses and quantization models (LOM, EPR). The test was inconclusive in estimating the agreement of coupling strengths between qubits and resonators, but it was excellent in estimating transition frequencies and anharmonicities.

The measured coherence times ($T_1$ and $T_2$) were lower than anticipated for both transmons. This discrepancy is attributed to non-idealities in the fabrication process and potential issues with the experimental setup, as discussed in Sec. \ref{sec:experiment}. To enhance coherence times, future designs will focus on reducing the surface participation ratio and investigating alternative qubit geometries. Further research will also aim to optimize the design strategy to enhance coherence times while maintaining substantial dispersive coupling, and to upscale the design by leveraging qubit-qubit interactions to investigate nonstandard photon detection scenarios for dark matter search applications. Another important future direction involves engineering the detection setup, including its coupling with the storage cavity.
\begin{acknowledgments}
This work was supported by Qub-IT, a project funded by the Italian Institute of Nuclear Physics (INFN) within the Technological and Interdisciplinary Research Commission (CSN5), by PNRR MUR projects PE0000023-NQSTI and CN00000013-ICSC, and by PRIN MUR project IRONMOON 2022BPJL2L. The authors would like to extend their thanks to S. Benz, J. Bieseker, and A. Sirois for their insightful discussions. \\
A substantial part of the methodologies and results described in this manuscript was presented at the IEEE Workshop on Low-Temperature Electronics, 16\textsuperscript{th} edition - WOLTE16.
\end{acknowledgments}

\bibliography{bibliography.bib}

\end{document}